\documentclass[preprint,12pt, a4paper]{elsarticle}

\usepackage{amssymb}
\usepackage{amsmath}
\usepackage{lineno}

\usepackage{float}
\restylefloat{table}
\usepackage{caption}

\journal{SoftwareX}

\usepackage{xcolor}

\begin{document}

\begin{frontmatter}



\title{\emph{qNoise}: A generator of non-Gaussian colored noise}

\author[a,b]{J. Ignacio Deza\corref{author}} \author[a]{Hisham Ihshaish}

\cortext[author] {Corresponding author.\\\textit{E-mail address:} ignacio.deza@uwe.ac.uk}
%\address[a]{DONLL, Universitat Polit\`ecnica de Catalunya, C/ Jordi Girona 31, 08034, Barcelona, Spain}
\address[a]{Faculty of Environment and Technology,\\ University of the West of England, Bristol, BS16 1QY, United Kingdom}
\address[b]{Universidad Atl\'antida Argentina, Arenales 2740, Mar del Plata, B7600EGN, Argentina}

\begin{abstract}
We introduce a software generator for a class of \emph{colored} (self-correlated) and \emph{non-Gaussian} noise, whose statistics and spectrum depend on two parameters, $q$ and $\tau$. Inspired by Tsallis' nonextensive formulation of statistical physics, the so-called $q$-distribution is a handy source of self-correlated noise for a large range of applications. The $q$-noise---which tends smoothly for $q=1$ to Ornstein--Uhlenbeck noise with autocorrelation $\tau$---is generated via a stochastic differential equation, using the Heun method (a second order Runge--Kutta type integration scheme). 
The  algorithm is implemented as a stand-alone library in \texttt{C++}, and is made available as open source in the \texttt{Github} repository. Noise' statistics can be specified handily; by only varying parameter $q$: it has compact support for $q<1$ (sub-Gaussian regime) and finite variance up to $q=5/3$ (supra-Gaussian regime). Once $q$ is fixed, noise' autocorrelation can be tuned independently by means of parameter $\tau$. 
The presented \emph{qNoise} generator provides a readily tool to modeling wide range of real-world noise types, and is suitable to study the effects of correlation and deviations from the normal distribution in systems of stochastic differential equations, key to understand system dynamics in numerous applications. The effect of noises' statistics on the response of a range of nonlinear systems is briefly discussed. In many of these examples, the systems' response turns optimal for some $q\neq1$. 
Hence, this paper aims to introduce \emph{qNoise} generator for \texttt{C++} at the class level and evaluate the kind of noise it generates, alongside their use in a range of applications. 
\end{abstract}

\begin{keyword}
non-Gaussian \sep random process generator \sep stochastic differential equations
\end{keyword}

\end{frontmatter}

%\linenumbers

% Computer program descriptions should contain the following
% PROGRAM SUMMARY.

%%===============================
\section{Motivation}\label{sec:1}
%%===============================

Most studies on noise-induced phenomena \cite{gosa99,ssgo07} have assumed the noise source to have Gaussian distribution, either ``white" (memoryless) or ``colored" (red, pink, etc \ldots). Although customarily accepted without criticism on the basis of the central limit theorem, the true rationale behind this assumption lies in the possibility of obtaining some analytical results, and avoiding many difficulties arising in generating and handling non-Gaussian noise. There is however experimental evidence that at least in some cases (particularly in sensory and biological systems) non-Gaussian noise sources may add desirable features to noise-induced phenomena (e.g.\ robustness, fault tolerance \cite{ckfw01}). These findings add practical interest to the task of finding viable ways to deal with non-Gaussian noise.

Here we introduce a lightweight (generic \texttt{C++} class) generator for non-Gaussian, colored stochastic processes. The expected applications of this algorithm are as diverse as the modeling of some types of vibration or fluctuation which are typically non-Gaussian, the generation of noise which is naturally confined to a domain, or the investigation of the response of many dynamical systems embedded in noise, as the latter deviates from being Gaussian.

The main features of noise obeying Tsallis' statistics are summarized in Sec.\ \ref{sec:2}; section \ref{sec:3} is devoted to the description of the software architecture and properties; section \ref{sec:4}, provides statistical analysis of the generated noise in the qualitatively different cases, alongside a discussion on the $q$-dependence of the variance and effective self-correlation time. The measured self-correlation times of the obtained series are compared with a fitting expression \cite{fuwt02}. In Sec.\ \ref{sec:5}, a brief review is provided on related work, namely where non-Gaussian noise-induced phenomena have been studied. 

\section{$q$-noise with Tsallis' statistics}\label{sec:2}

The exponentially self-correlated Gaussian noise $\eta(t)$ named after Ornstein and Uhlenbeck (OU noise, or ``colored'' Gaussian noise) can be \emph{dynamically} generated through the differential equation
\begin{equation}\label{eq:OU}
\tau\,\dot{\eta}=-\eta(t)+\xi(t),
\end{equation}
where $\xi(t)$ is centered Gaussian white noise with variance $D$, namely 
\[\langle\xi(t)\rangle=0,\quad\langle\xi(t)\,\xi(t')\rangle=2D\,\delta(t-t').\]
This way, $\eta$'s self-correlation time is $\tau$.

A straightforward generalization of Eq.\ (\ref{eq:OU}) was proposed by Borland some time ago \cite{libo98} as a model for \emph{correlated} diffusion:
\begin{equation}\label{eq:qN}
\tau\,\dot{\eta}=-\frac{d}{d\eta}V_{q}(\eta)+\xi(t)
\end{equation}
where the potential $V_q$ is given by:
\begin{equation}\label{eq:Vq}
 V_{q}(\eta)=\frac{D}{\tau\,(q-1)}\,\ln\left[1+\frac{\tau\,(q-1)}{D}\,\frac{\eta^2}{2}\right],
\end{equation}
As much as the OU noise allows to explore spectral effects within the class of exponentially correlated noise, this generalization provides moreover a device to explore statistics effects by varying just one parameter (namely $q$, at constant $\tau$ and $D$). 

The stationary properties of $\eta$ (including its autocorrelation function) are thoroughly described elsewhere \cite{fuwt02,hwio05,hwio04,libro,wide13}, we here summarize the main results. Using the Fokker--Planck formalism, one obtains the stationary probability distribution
\begin{equation}\label{eq:pdf}
P_q^{\mathrm{st}}(\eta)=\frac{1}{Z_q}\left[1+\frac{\tau\,(q-1)}{D}\,\frac{\eta^2}{2}\right]^{\frac{1}{1-q}},
\end{equation}
which can be normalized only for $q<3$ ($Z_q$ is a normalization factor). The first moment $\langle\eta\rangle$ always vanishes \cite{fuwt02,hwio05,hwio04,libro,wide13} and the second moment,
\begin{equation}\label{eq:2nd}
\langle\eta^2\rangle=\frac{2D}{\tau\,(5-3q)},
\end{equation}
is finite only for $q<5/3$.

\begin{figure}[!ht]
\begin{center}
\includegraphics[width=0.48\textwidth]{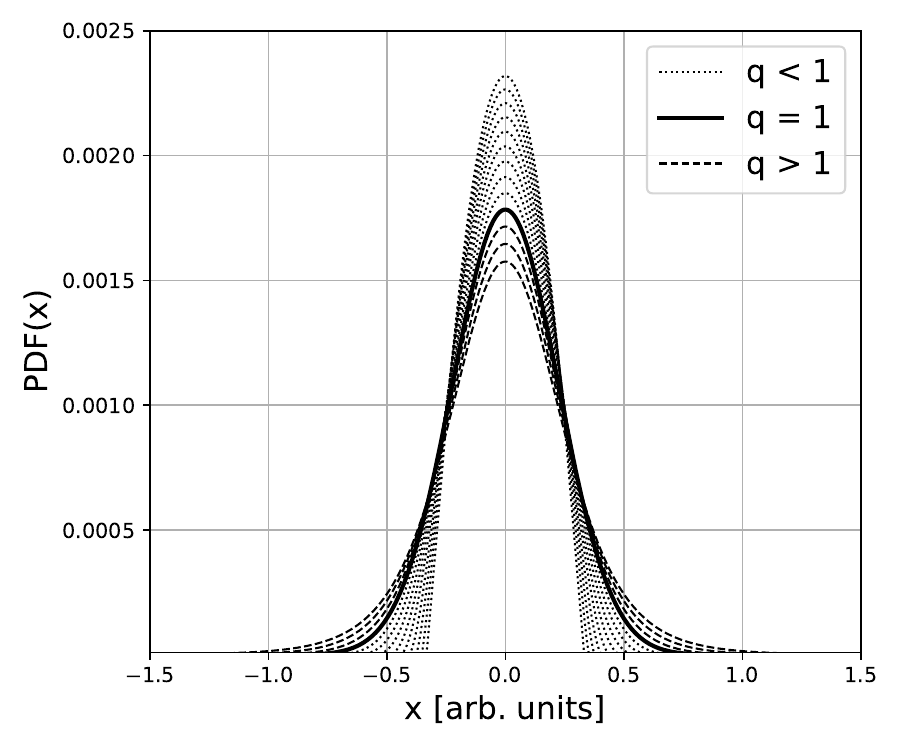}
\includegraphics[width=0.48\textwidth]{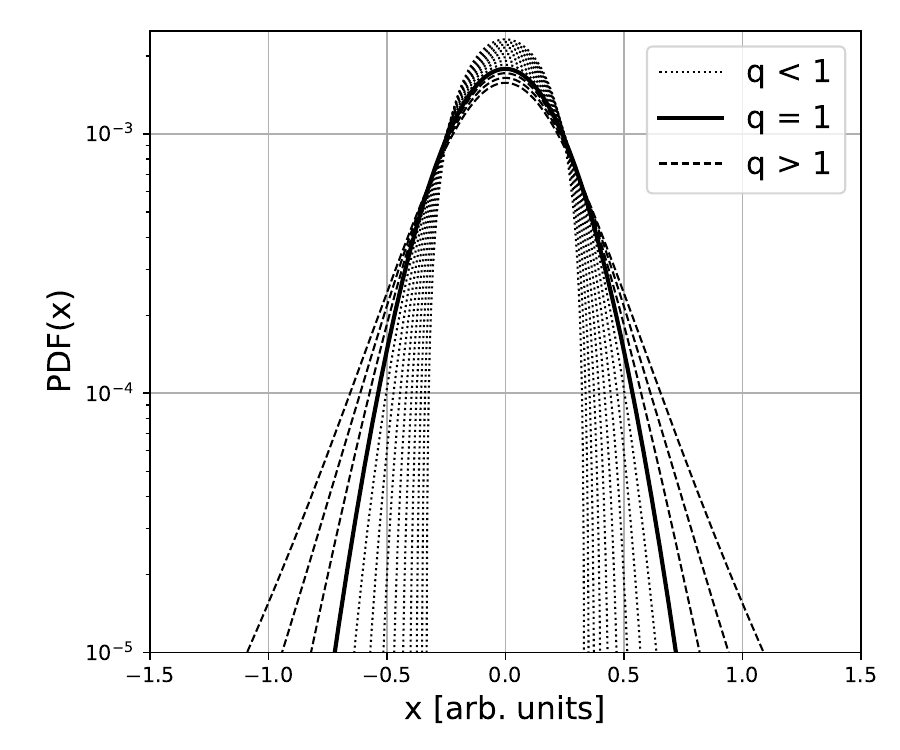}
\caption{Stationary $q$-noise pdf for $0<q<1$ (dotted line), $q=1$ (bold line) and $1<q<1.6$ (single line). The right panel show the same plot in semilogarithmic scale.  Notice the pdf is compact-supported for $q<1$, Gaussian for $q=1$ and fat-tailed for $q>1$.}\label{fig:1}
\end{center}
\end{figure}

Some properties of the noise are summarized in Fig.\ \ref{fig:1}. The bold line depicts the Gaussian limit ($q=1$). Curves of weaker full lines show that for $q>1$, the second moment is larger than the Gaussian limit $D/\tau$. For $q<1$ (dotted lines) the distribution has a cut-off and is only defined for 
 \begin{equation}\label{eq:cutOff}
|\eta|<\eta_c\equiv\sqrt{\frac{2D}{\tau\,(1-q)}}.
\end{equation}
Same distributions are shown in linear and semilogarithmic scales (Fig.\ \ref{fig:1}, left and right panels respectively).

The autocorrelation time $\tau_q$ of the process $\eta(t)$ in its stationary regime, also diverges for $\to5/3 \approx 1.66$. Far from its divergence point, it can be approximated as in \cite{fuwt02}:
\begin{equation}\label{eq:tauDiv}
\tau_q\approx\frac{2\tau}{5-3q}.
\end{equation}

When $q\to1$,  $\eta$ becomes a Gaussian colored noise, namely the Ornstein--Uhlenbeck process $\xi_{OU}(t)$, with correlations
\begin{equation}\label{eq:OUcorr}
\langle\xi_{OU}(t)\,\xi_{OU}(t')\rangle=\frac{D}{\tau}\exp\left(-\frac{|t-t'|}{\tau}\right),
\end{equation}
and probability distribution
\begin{equation}\label{eq:OUPDF}
P^{\mathrm{st}}(\xi_{OU})=Z^{-1}\exp\left(-\frac{\tau}{D}\,\frac{\xi_{OU}^2}{2}\right).
\end{equation}

\subsection{ $\tau$ and $\langle\eta^2\rangle$  dependence on $q$}\label{sec:tauq}

\begin{figure}[!h]
\begin{center}
\includegraphics[width=.48\textwidth]{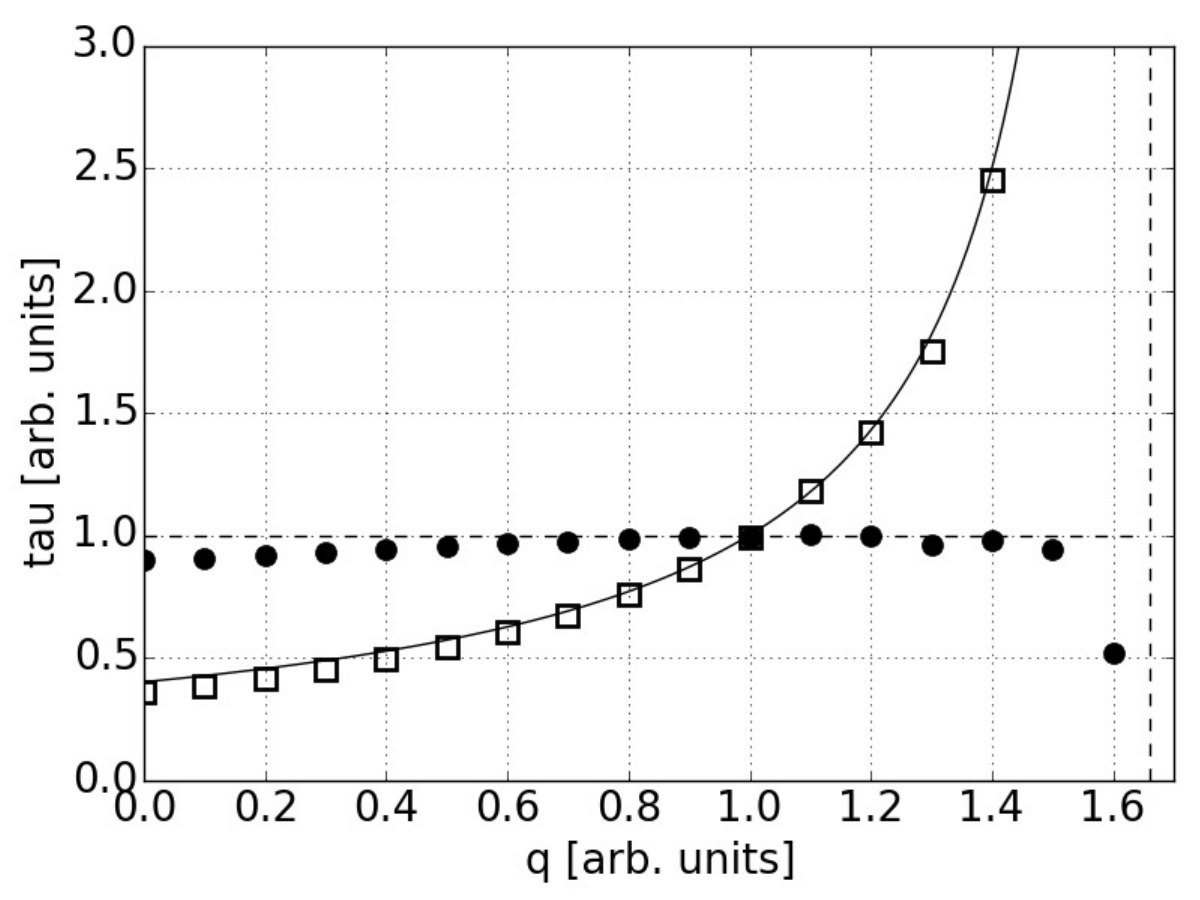}
\includegraphics[width=.48\textwidth]{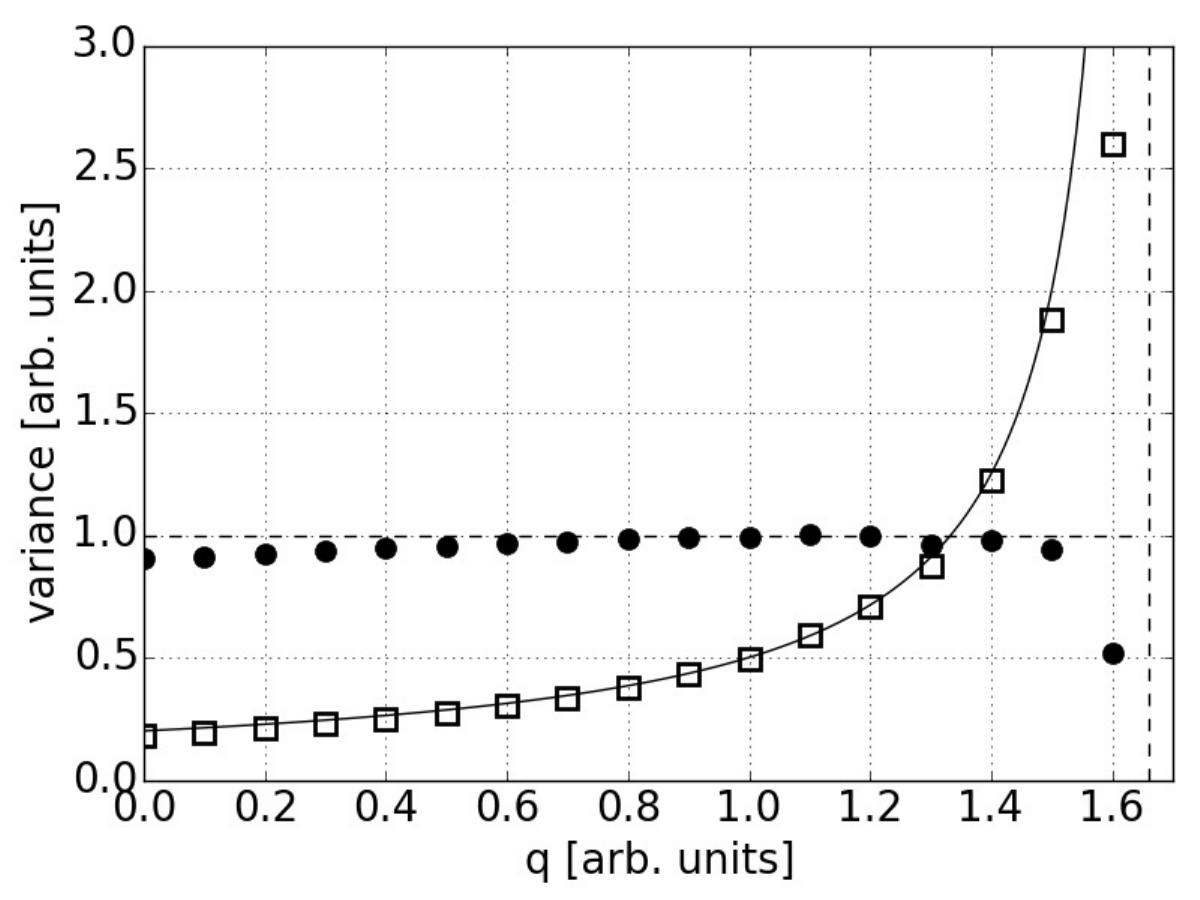}
\caption{(Left) Dependence of $\tau$ on $q$. The white squares show the measured dependence, fitted by Eq.\ (\ref{eq:tauDiv}) (full line). The black dots show the behavior of an effective $\tau$ in order to make it independent of $q$. (Right) Equivalently for the variance of the noise, using Eq.\ (\ref{eq:2nd}).  }\label{fig:effective}
\end{center}
\end{figure}

Equations Eqs.\ (\ref{eq:2nd}) and (\ref{eq:tauDiv}) tell us that for $q\ne1$, $\langle\eta^2\rangle$ and $\tau_q$ do not attain their values ($D$ and $\tau$ respectively) in a normal distribution. Rather, they both diverge at  $q=5/3$ (white squares in both panels of Fig.\ \ref{fig:effective}). It is however desirable to have a generator able to approximately keep constant the characteristics of these properties with respect of $q$, at least sufficiently far away from the divergence point. This can be very useful to study the effects of the statistics due to changes in $q$ keeping  $\tau$ and variance constant. 
 
This way an \emph{effective} $\tau_q$ and $\langle\eta^2\rangle$ can be defined by dividing $\tau$ by Eq.\ (\ref{eq:tauDiv}) before integration and $\langle\eta^2\rangle$ by Eq.\ (\ref{eq:2nd}) after integration. The filled circles in both panels of Figure \ref{fig:effective} show this dependence for both $\tau$ and $\langle\eta^2\rangle$, and how the system becomes independent of $q$ for  the range $0<q<1.5$ approximately. For $q>1.5$, the  proximity to the divergence point $q=5/3$ (shown with a dotted line) makes this approximation fail.

\section{Generator \texttt{<class>} Description}\label{sec:3}

The noise generator \cite{qNoise} is implemented in \texttt{C++} as class, with dependencies on standard libraries only. It generates random numbers using functions in the built-in \texttt{<random>} class.
 The generator provides functions for Gaussian white noise, Gaussian colored noise (Orstein-Uhlenbeck), and two versions of non-Gaussian non-white noise. One where $\tau$ and $\langle\eta^2\rangle$  \emph{depend} on $q$ (as in Eqs. \ref{eq:2nd} and \ref{eq:tauDiv}) and a normalized version where this effect has been counterbalanced to the first order, sufficiently far away from $q=5/3$ (as shown as black dots in figure \ref{fig:effective}). This batch of functions would facilitate modeling a wide variety of scenarios and is suitable for many applications, some of which are detailed in the last section of this paper.
 By default, the tool uses the Mersenne-Twister generator \cite{mersenne} which provides a very long ($2^{19937} - 1$) pseudo-random number cycle. Hence it is advised to seed the generator \emph{only once} to avoid spurious correlations.

\subsection{Functions}
\begin{figure}[htb]
\begin{center}
\includegraphics[width=.5\textwidth]{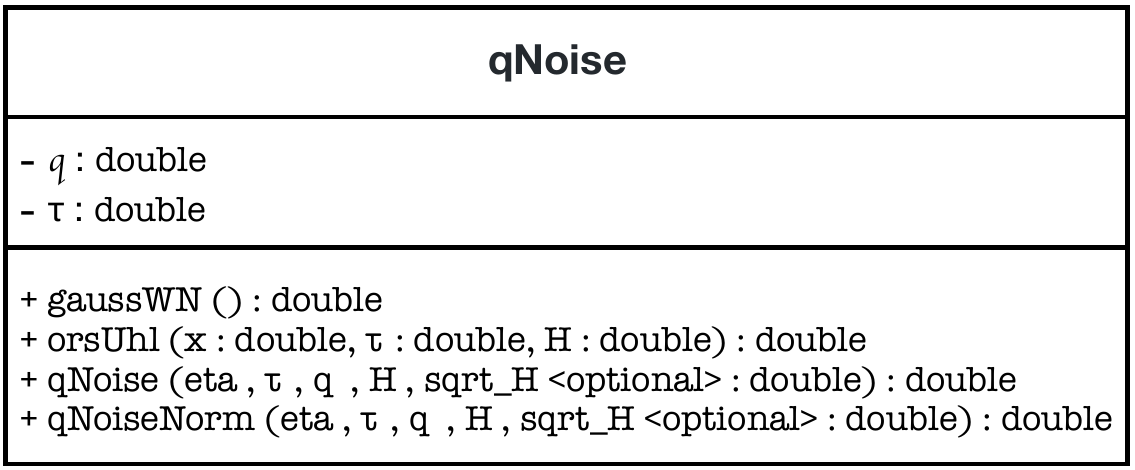}
\caption{Class diagram of \emph{qNoise} illustrating the four implemented functions.}
\label{fig:class}
\end{center}
\end{figure}

The class implements four public member functions as shown in Fig. \ref{fig:class}.   

%\begin{small}
%\begin{verbatim}
%1. double gaussWN ()
%2. double orsUhl (double x, double tau, double H)
%3. double qNoise (double eta, double tau, double q, 
%		double H, double sqrt_H (optional))		
%4. double qNoiseNorm (double eta, double tau, double q, 
%		double H, double sqrt_H (optional))
%\end{verbatim}
%\end{small}

The first function is a wrapper for the normal distribution, implemented in the \texttt{<random>} standard library. It is presented as a function of this class for convenience. 

The second function is an implementation of the Orstein-Uhlenbeck noise. It accepts three parameters. The previous value of the noise (since it is a Markov process),  the autocorrelation time $\tau$ of the noise and the integration time $H$ (necessary for setting the adequate timescale of the noise).

The third function implements the q-noise distribution. It accepts the same variables as the \texttt{orsUhl} function in addition to $q$ (the noise statistics), and sqrt\_H as an optional variable. If $H$ is constant, explicitly setting \texttt{sqrt\_H} = $H^{1/2}$  will avoid its calculation every time the function is called. A snippet of the function is shown in Listing 1 below.   

Finally, the fourth function is a wrapper for the third function. Here $\tau$ is given  by Equation (\ref{eq:tauDiv}) and the resulting noise is divided using Eq. (\ref{eq:2nd}) in order to counterbalance the dependence of both $\tau$ and the variance of the noise on $q$. See section \ref{sec:tauq} for an analysis of this effect and a discussion about its range of validity.

\begin{figure}[h]

\includegraphics[width=.45\textwidth]{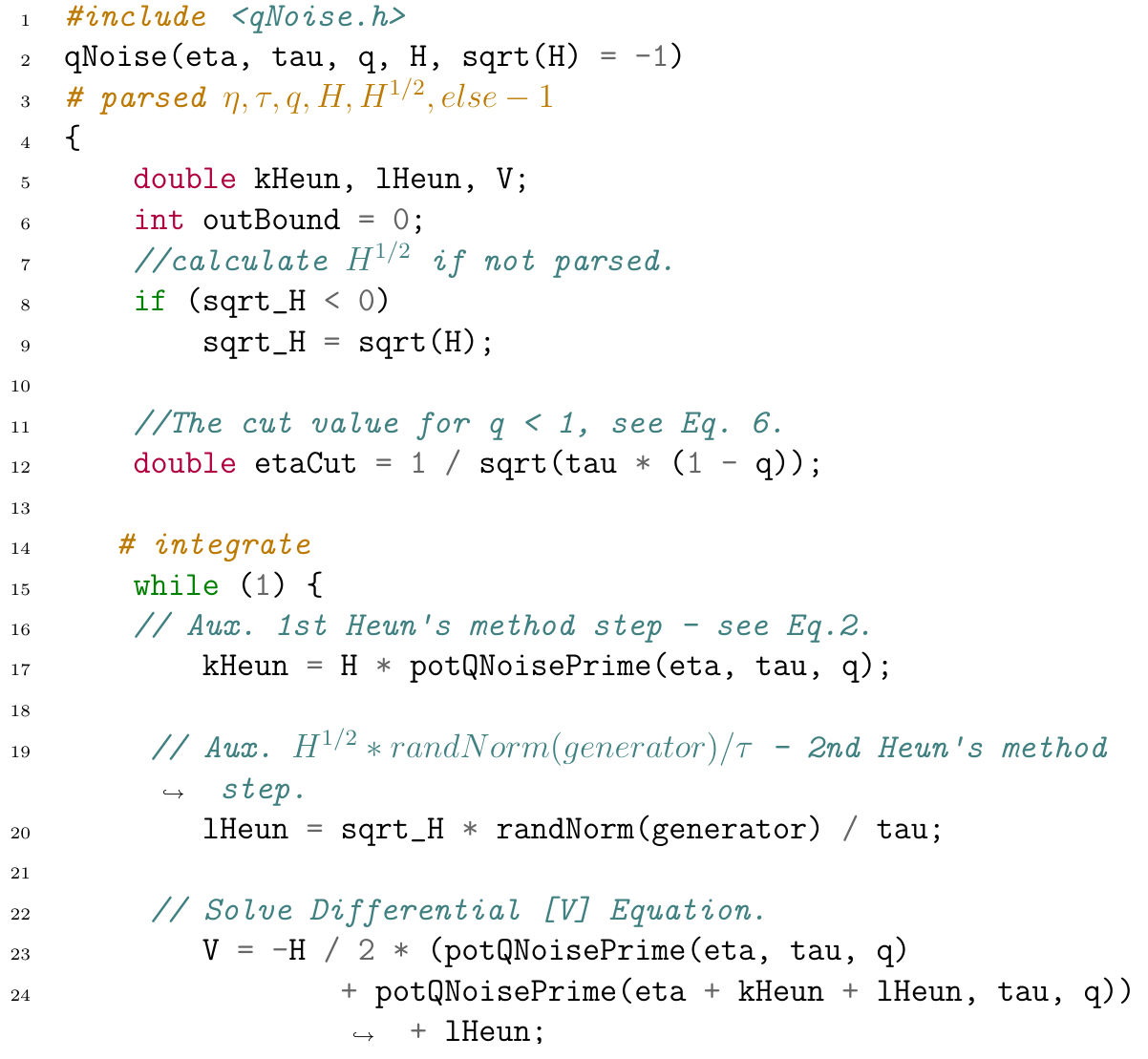}
\vspace{2cm}
\includegraphics[width=.5\textwidth]{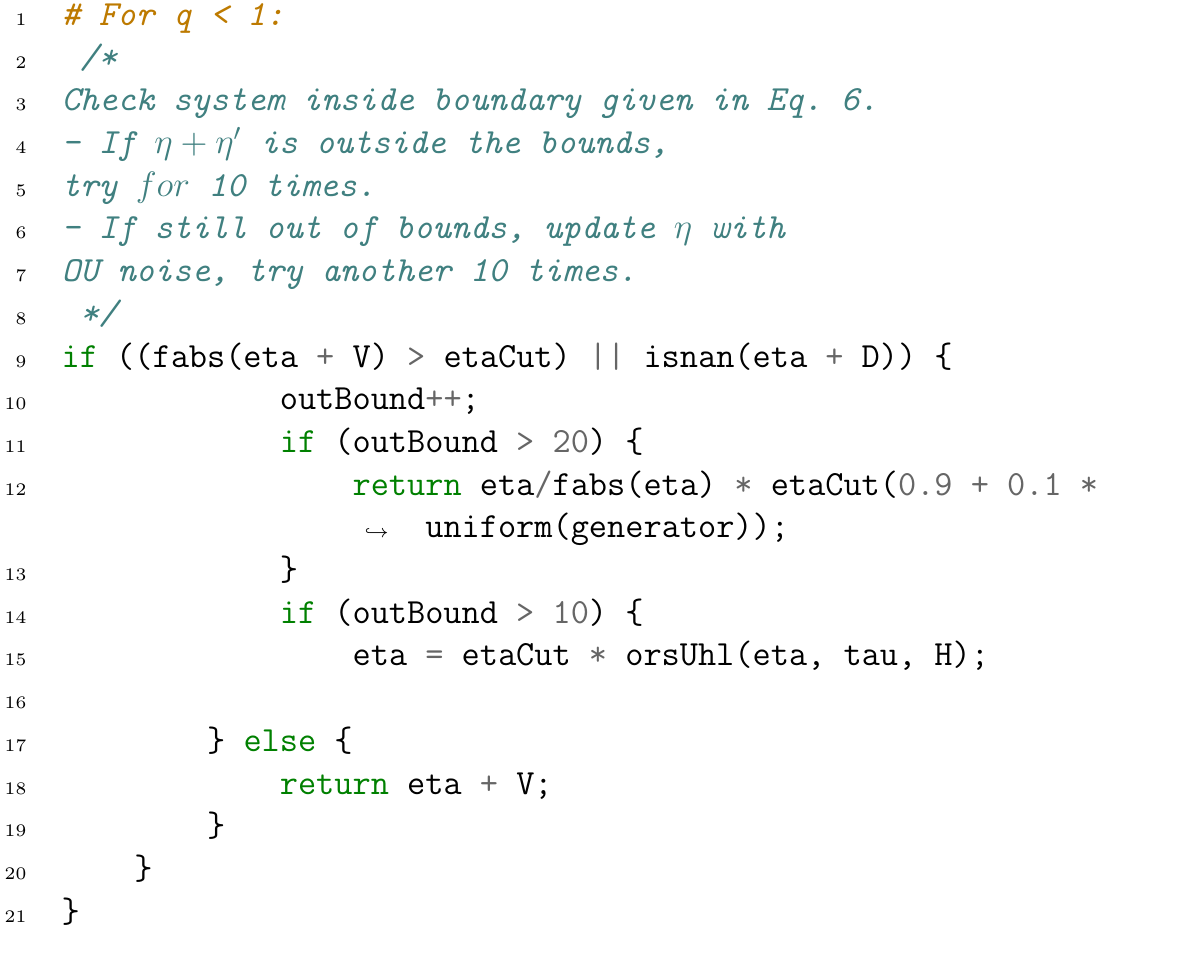}
{\caption*{Listing 1: \small{\texttt{qNoise()} pseudocode: integrates a differential equation using the Heun Method. For $q=1$, returned noise behaves like OU noise, for $q<1$, noise not defined outside (+/- etaCut) and for $q>1$, noise statistics become supra-Gaussian. Code and documentation are available in \cite{qNoise}.}}}
\label{lst:qnoise}
\end{figure}

\begin{figure}[!ht]
\begin{center}
\includegraphics[width=.8\textwidth]{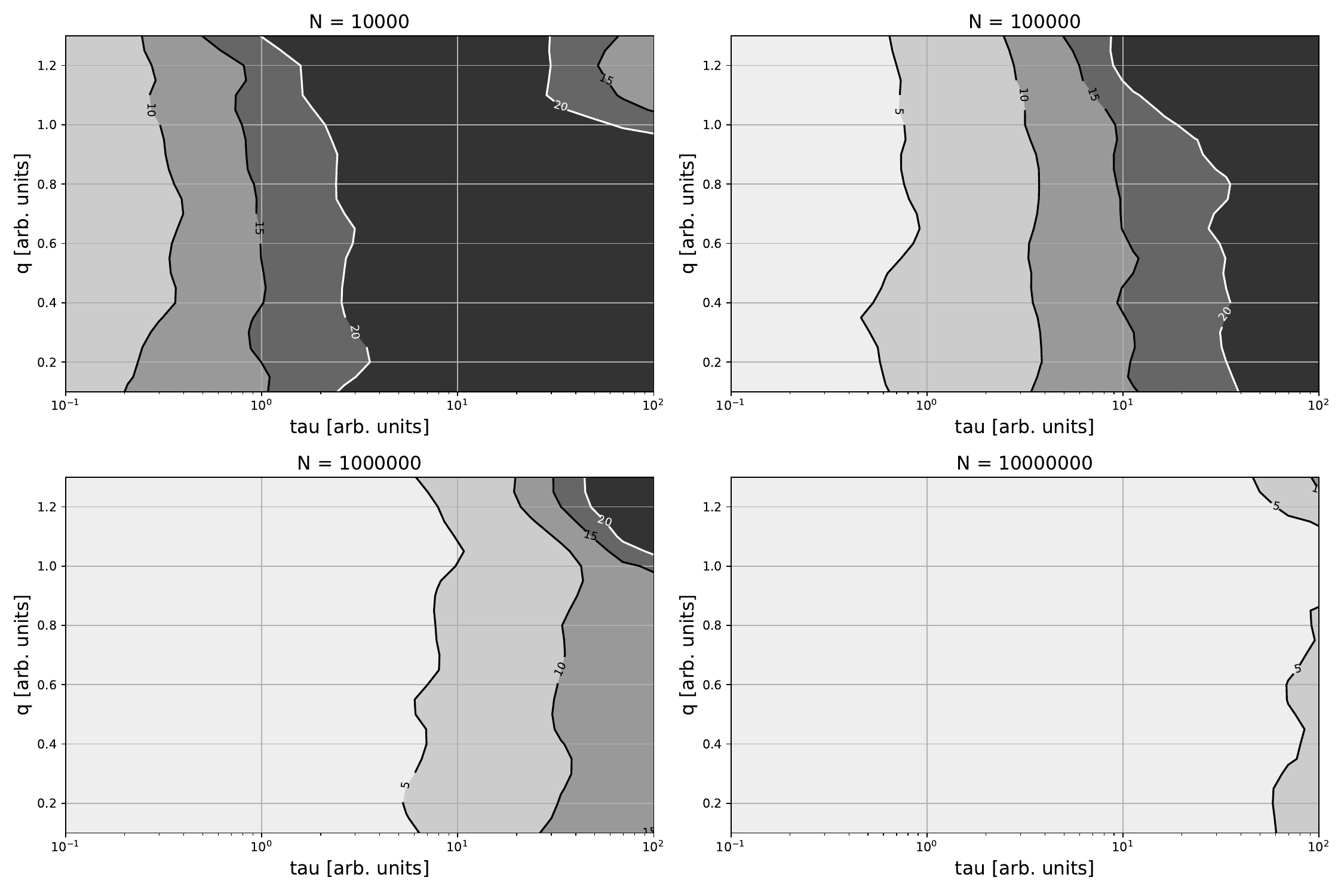}
\caption{Accuracy of the histogram of the noise (calculated as the square root of square's sum of the difference of the generated noise' histogram and the theoretical distribution $\sqrt{\sum_{x} (h(x) - \text{pdf}(x))^2}$). This shows the accuracy of the generated noise — note its dependency of $N$ and its relative independence of $q$.}
\label{fig:unittest}
\end{center}
\end{figure}

A generic unit test results are shown in Fig. \ref{fig:unittest}. The test compares the generated average noise, of an ensemble of 10 \texttt{qNoise} runs for each set of $q$, $\tau$ and $N$, with the expected distribution of noise. As expected, only when $N$ is relatively small does the generated noise deviate from its theoretical distribution, particularly for high $\tau$. That is, it takes longer (higher $N$) for a highly correlated noise (high $\tau$) to explore the support and approximate the PDF.

\subsection{Seeding}
The presented class enables seeding the random number generator in two functions:
\begin{figure}[h]
\includegraphics[width=.45\textwidth]{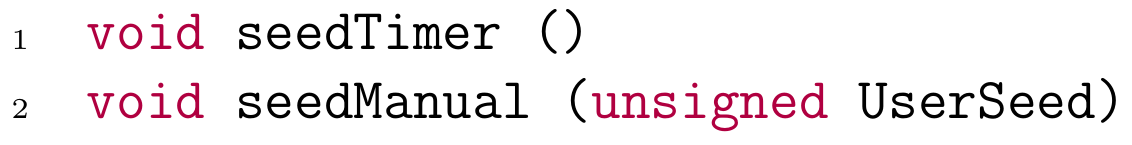}
{\caption*{}}
\label{lst:qnoise}
\end{figure}

Timer seeding is provided as the default setting for random number generator, and it is done automatically. For lightweight single-threaded runs, manual seeding is not required. However, in multi-threading settings manual seeding (call \texttt{seedManual}) for each thread is recommended.

\section{Properties of the generated noise}\label{sec:4}

\subsection*{Bounded domain ($q<1$)}

\begin{figure}[!ht]
\begin{center}
\includegraphics[width=.45\textwidth]{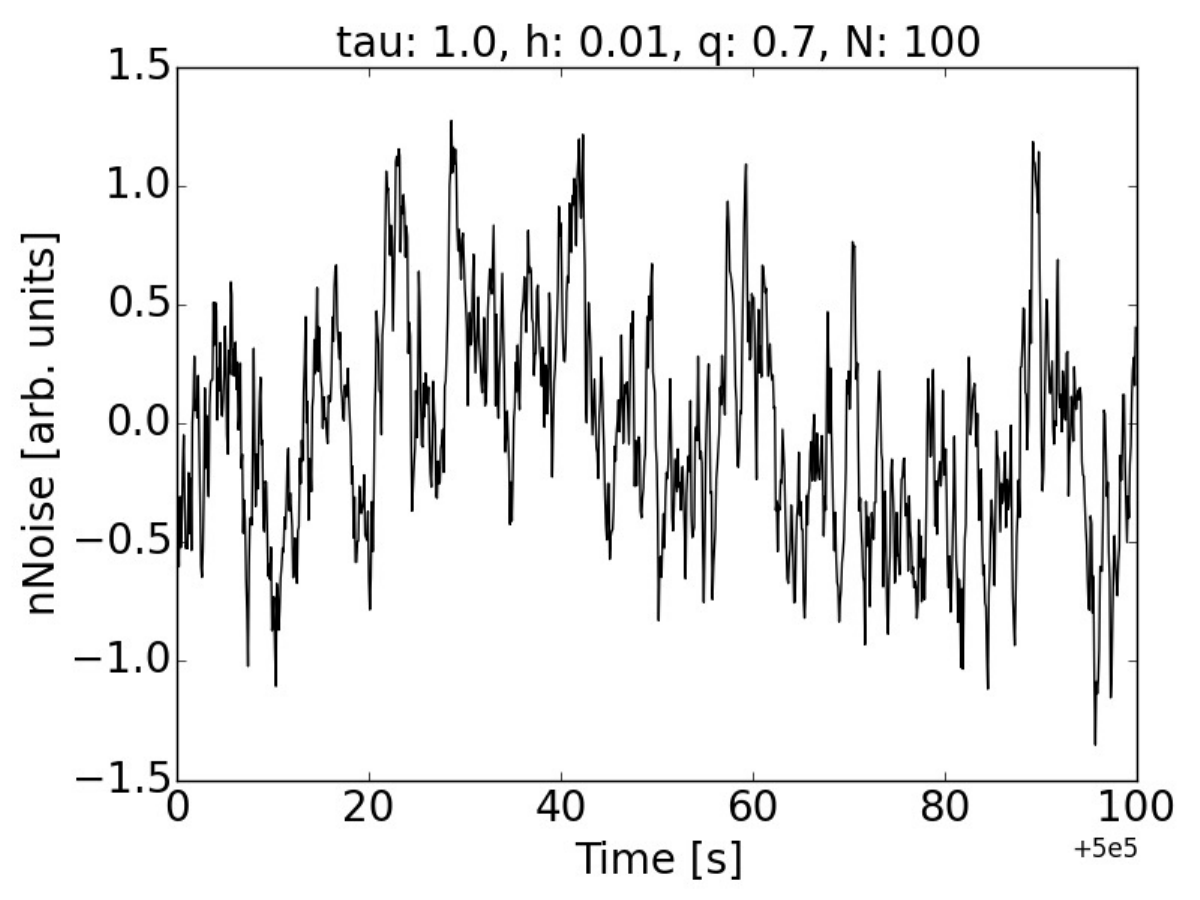}
\includegraphics[width=.45\textwidth]{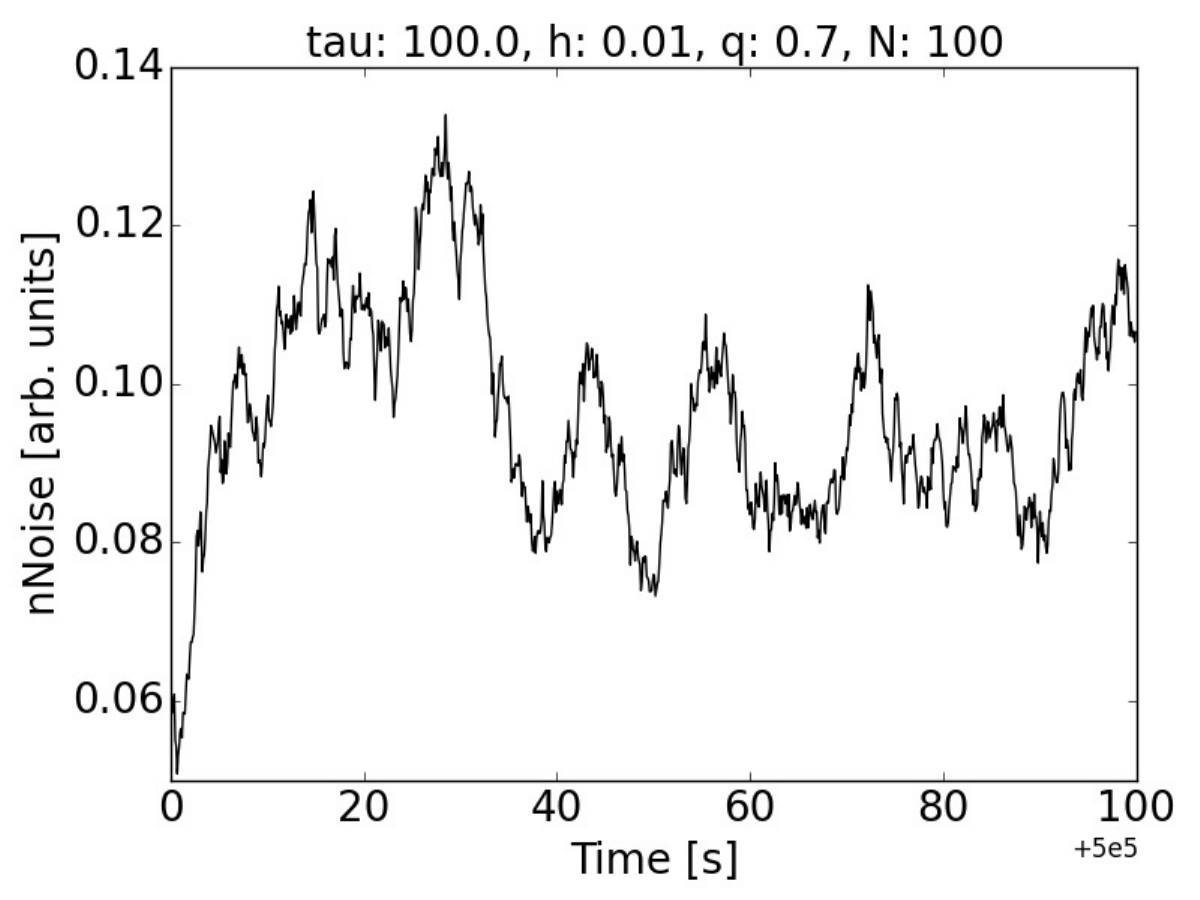}
\includegraphics[width=.45\textwidth]{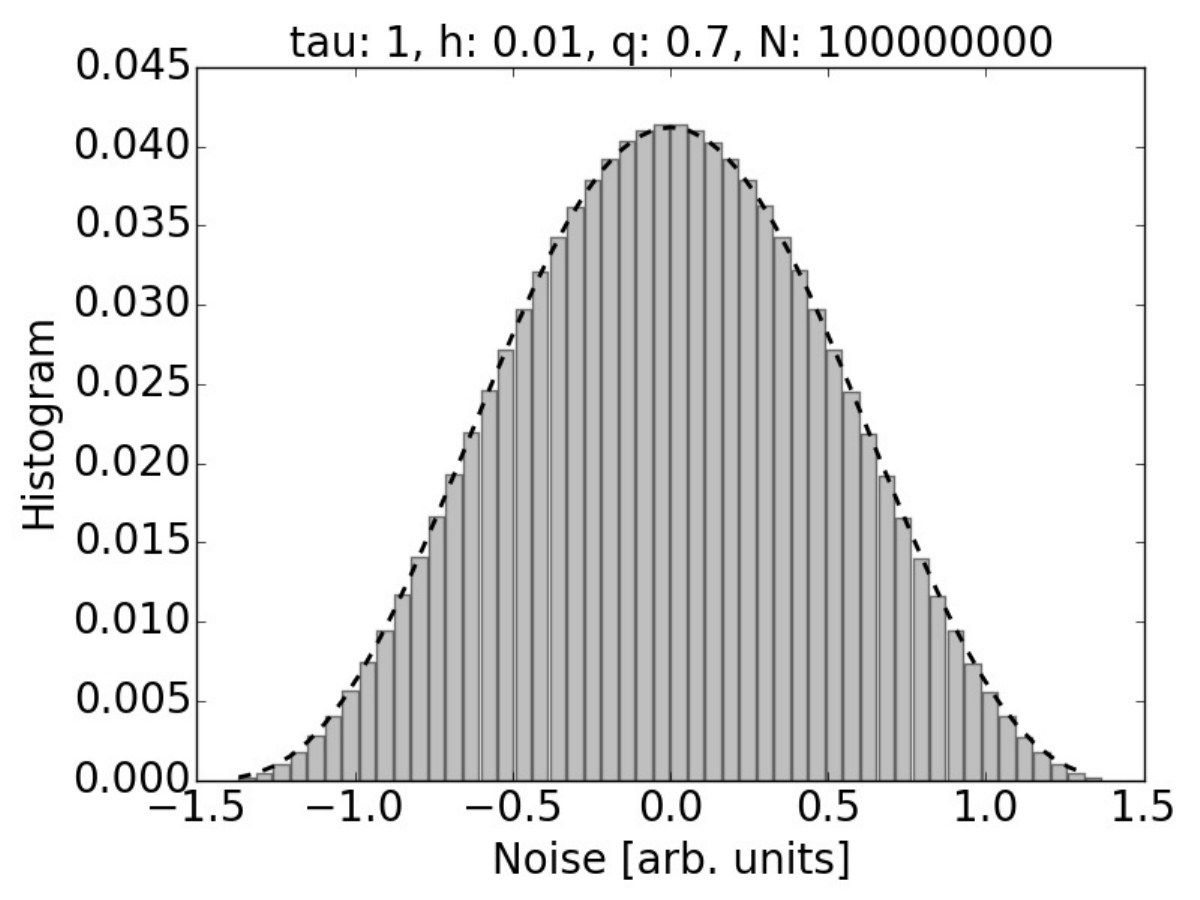}
\includegraphics[width=.45\textwidth]{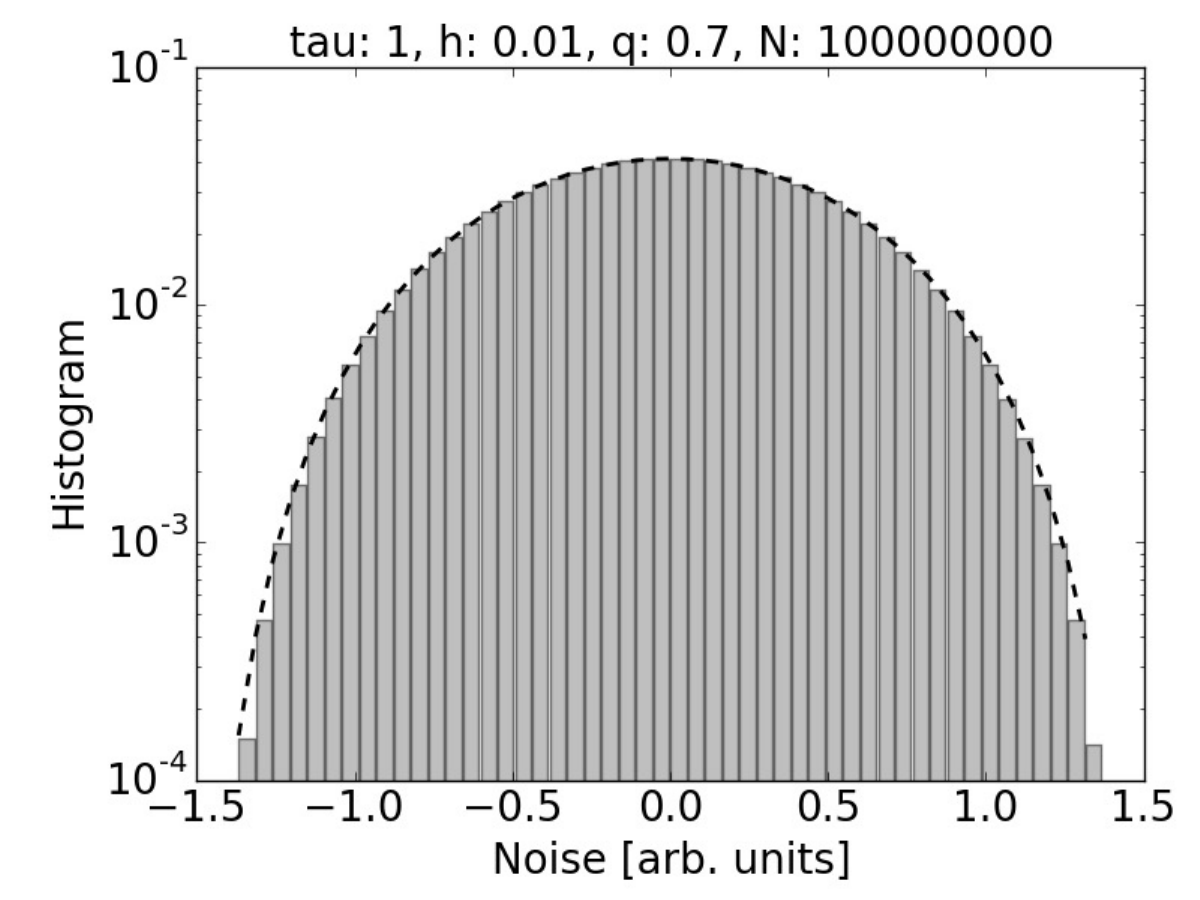}
\caption{q-noise for $q=0.7)$ (bounded domain),  and integration step $h=0.01$. The top panels show a sample of the generated noise, for (left) $\tau=1 s$ and (right) $\tau=100 s$. Notice that in the right figure the noise is not centered around zero as it is performing a very long excursion (larger than the sample) given its very high autocorrelation. Both histograms in the bottom panels show the same data, concurring with the sample on top-left. ($\tau=1 s$). Although in the linear histogram on the left it cannot be clearly seen, the semi-logarithmic plot on the right clearly show the bounded domain. The curve of dotted points shows the theoretical distribution as in Fig. \ref{fig:1} for the same parameters, which perfectly concurs with the histogram of the data. }\label{fig:qlessthan1}
\end{center}
\end{figure}

Bounded-domain noise is widespread in nature, and has multiple applications for modeling and control \cite{wide13}\footnote{In practice, physical noise has bounded domain because arbitrarily large fluctuations are strongly suppressed. Nonetheless, Gaussian noise has many desirable theoretic properties which allow for analytical results.}. The infra-Gaussian noise considered here can be addressed as a small deviation from Gaussianity, allowing a perturbative approach (Fig \ref{fig:qlessthan1}). In Sec. \ref{sec:resGate}, an example of a infra-Gaussian  noise is shown, in a resonant trap. Another use is as a source of noise whose distribution is quasi-normal but identically zero outside a boundary.

The noise generator algorithm does also ensure that noise domain is bounded, checking for out-of-bound values. This necessary test (especially for highly correlated noise) is implemented and documented accordingly in the provided source code.

\subsection*{Gaussian case ($q=1$)}

\begin{figure}[!ht]
\begin{center}
\includegraphics[width=.45\textwidth]{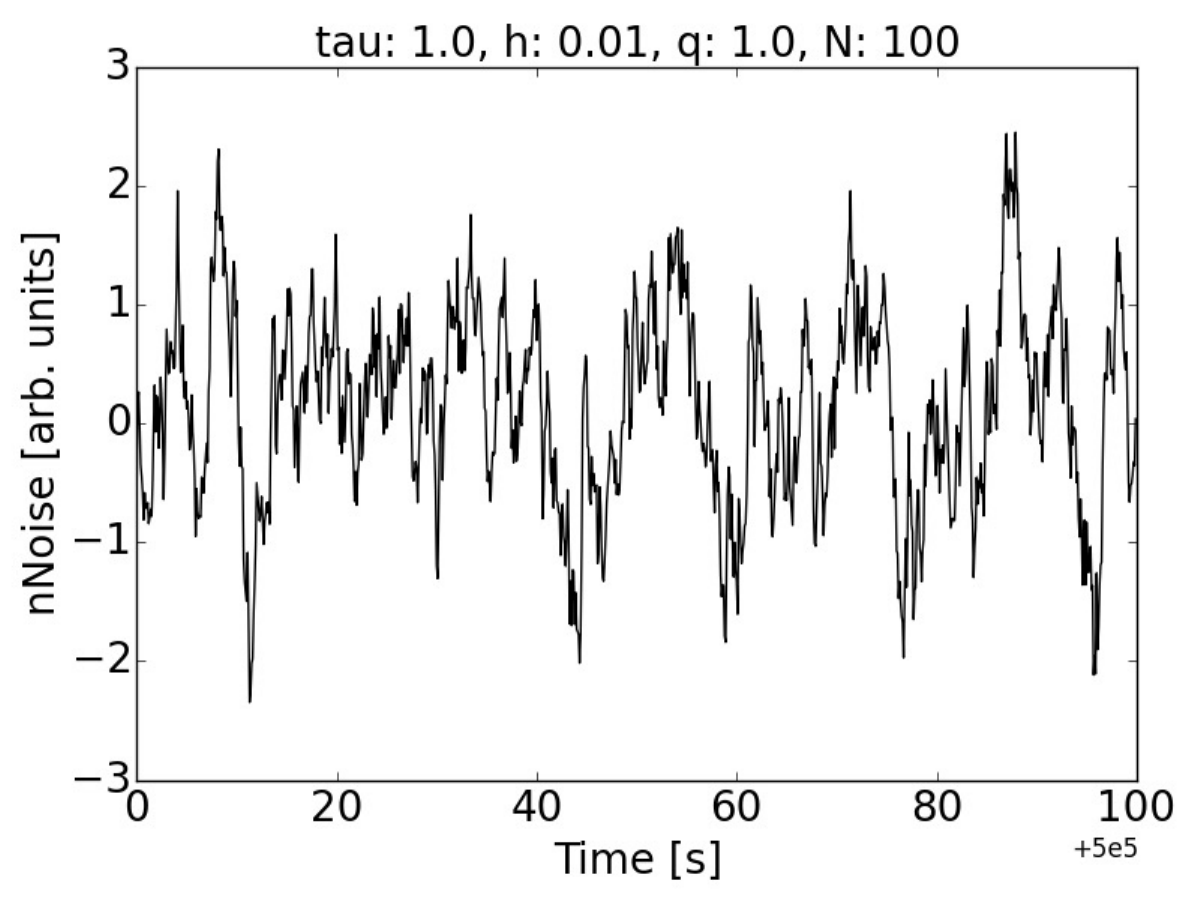}
\includegraphics[width=.45\textwidth]{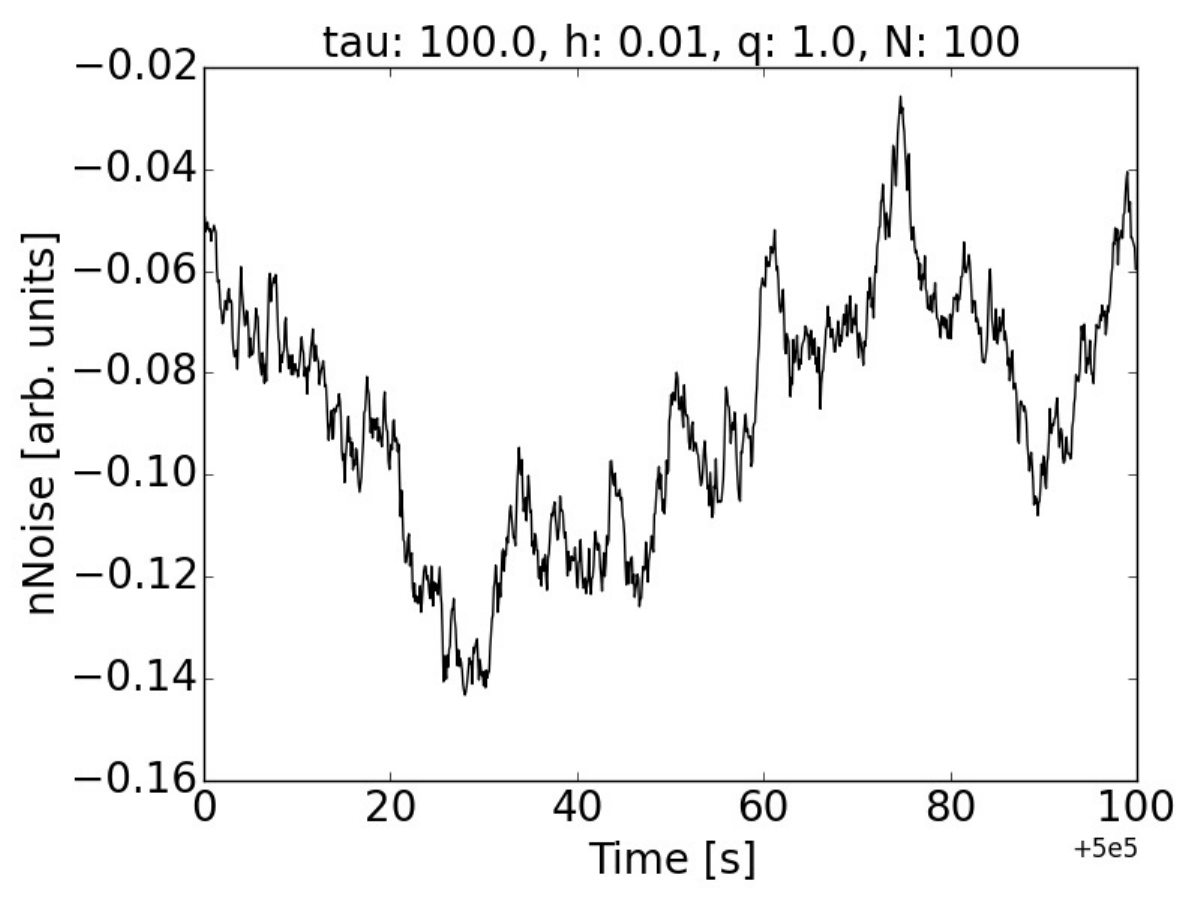}
\includegraphics[width=.45\textwidth]{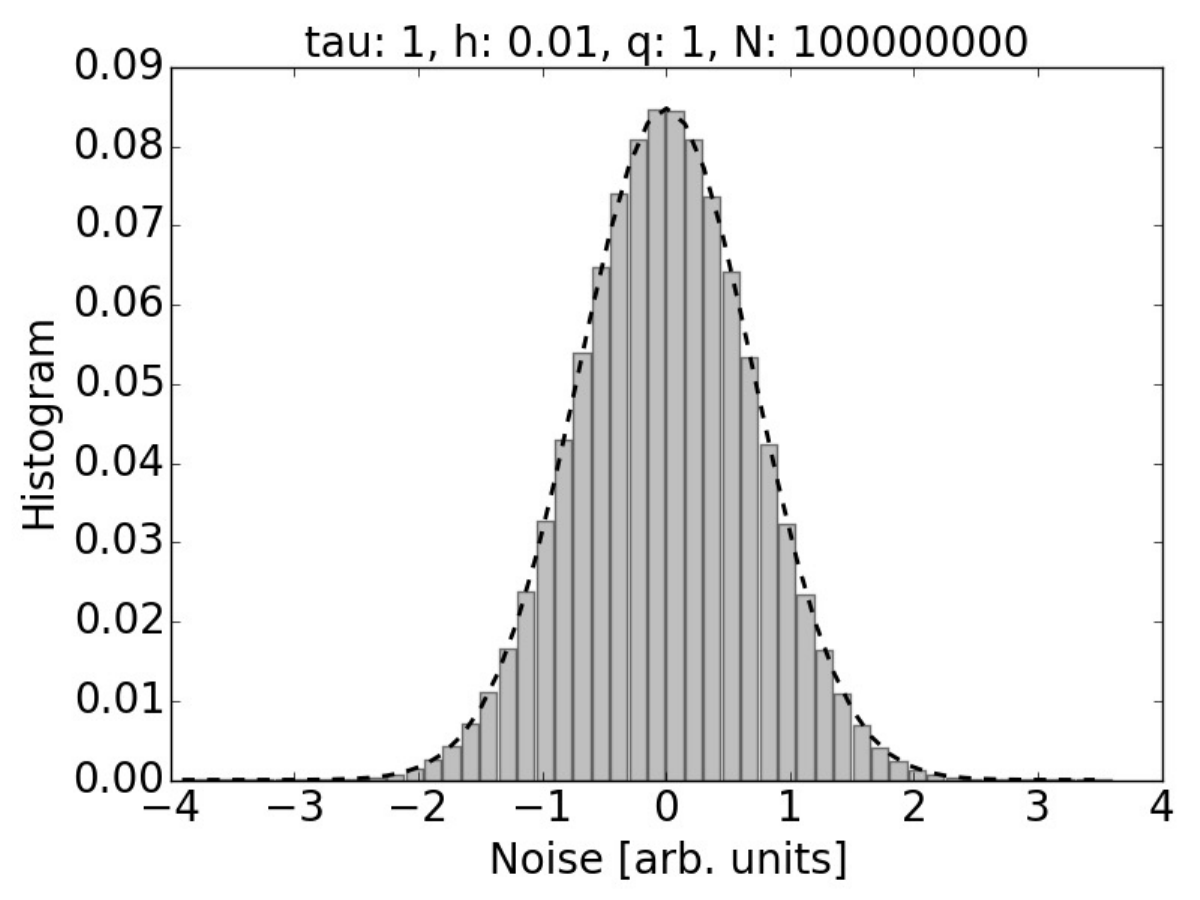}
\includegraphics[width=.45\textwidth]{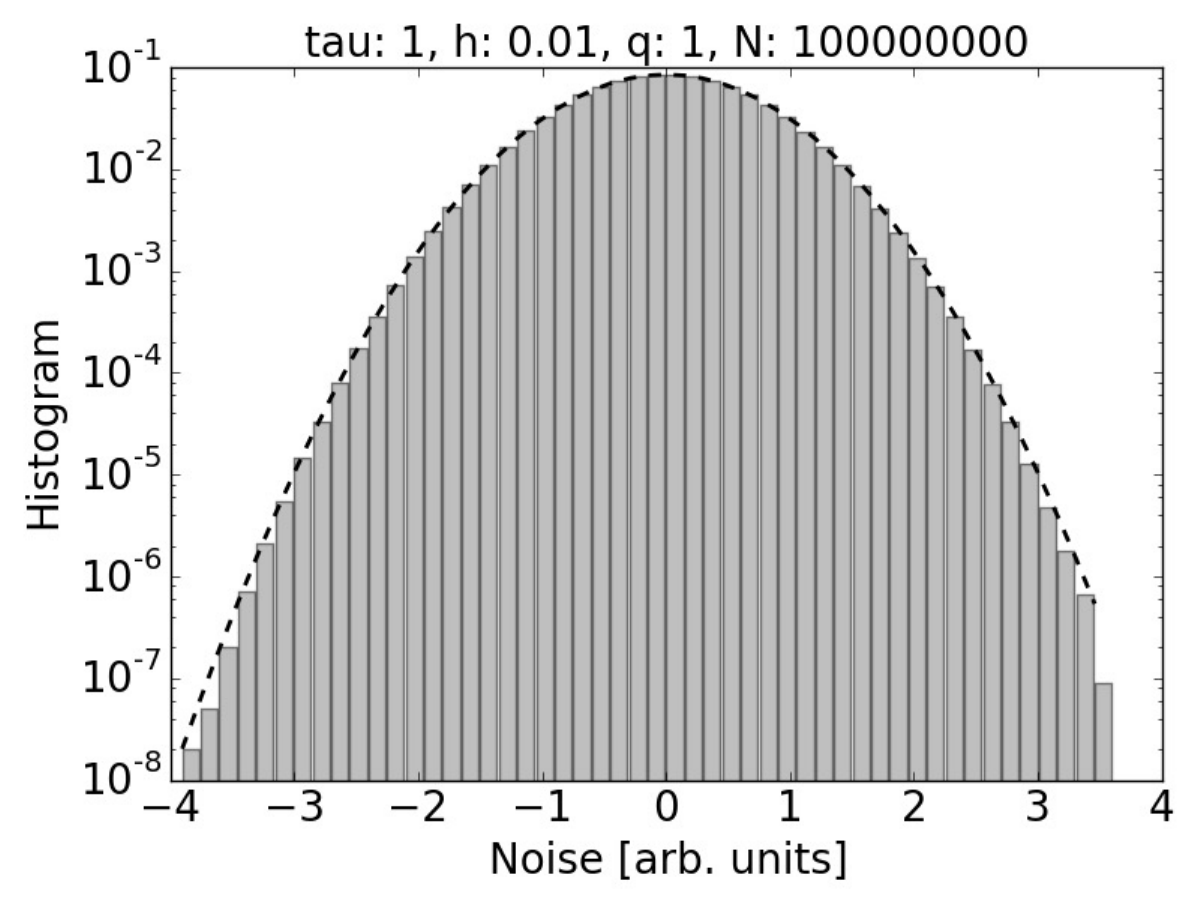}
\caption{q-noise for $q=1)$ (Gaussian Behavior),  and integration step $h=0.01$. The top panels show a sample of the generated noise, for (left) $\tau=1 s$ and (right) $\tau=100 s$. Notice that in the right figure the noise is not centered around zero as it is performing a very long excursion (larger than the sample) given its very high autocorrelation ($\tau$). Both histograms in the bottom panels show the same data, concurring with the sample on top-left. ($\tau=1 s$). Although the linear histogram on the left shows a bell-shaped distribution, this is not enough to demonstrate Gaussianity. However that is possible to observe on the semi-logarithmic, parabolic, plot on the right. The dotted points show the theoretical distribution as in Fig. \ref{fig:1} for the same parameters, which perfectly concurs with the histogram of the data. }\label{fig:qequalto1}
\end{center}
\end{figure}

The Gaussian case behaves exactly as an Orstein-Uhlenbeck noise, concurring perfectly with it for the whole range of $\tau$ (Fig \ref{fig:qequalto1}). As shown in the introduction, the limit $q\rightarrow1$ recovers the Gaussian noise, and all limits converge to it, see Eqs.  (\ref{eq:2nd})-(\ref{eq:OUPDF}). This limit allows to explore regions arbitrarily near the normal distribution. It can be used to model small deviations from it due to some underlying physical phenomenon. As the value of $q$ can be changed continuously and dynamically, this scheme also allows to model departures from the normal distribution due to long time-scale fluctuations, by slowly varying $1-\epsilon < q < 1 + \epsilon$ as a more realistic model for a small noisy system.

It is not generally recommended to compute the purely Gaussian case from the general case and set $q=1$, particularly due to potential computational complexity. An extensive batch of tests has been run in order to compare it to the Orstein-Uhlenbeck noise. All of the results were successfully recovered.
As presented above, \texttt{orsUhl}, a function for generating Orstein-Uhlenbeck noise for a variable $\tau$ is included in this package and its results are equivalent to using the non-normalized \texttt{qNoise} function for $q=1$ at a fraction of the computation time.

\subsection*{Supra-Gaussian noise ($1<q<5/3$)}

The supra-Gaussian (also called fat-tail) noise presented here is of the class of finite variance. This is usually an overlooked, modestly studied, class of noise. The Supra-Gaussian noise, generally considered in literature, tends to be \textit{L\'evy-like}, where the variance is infinite \footnote{The q-noise presents infinite variance for $q>5/3$ but the description of this behavior is outside the scope of this article.}.

\begin{figure}[!ht]
\begin{center}
\includegraphics[width=.45\textwidth]{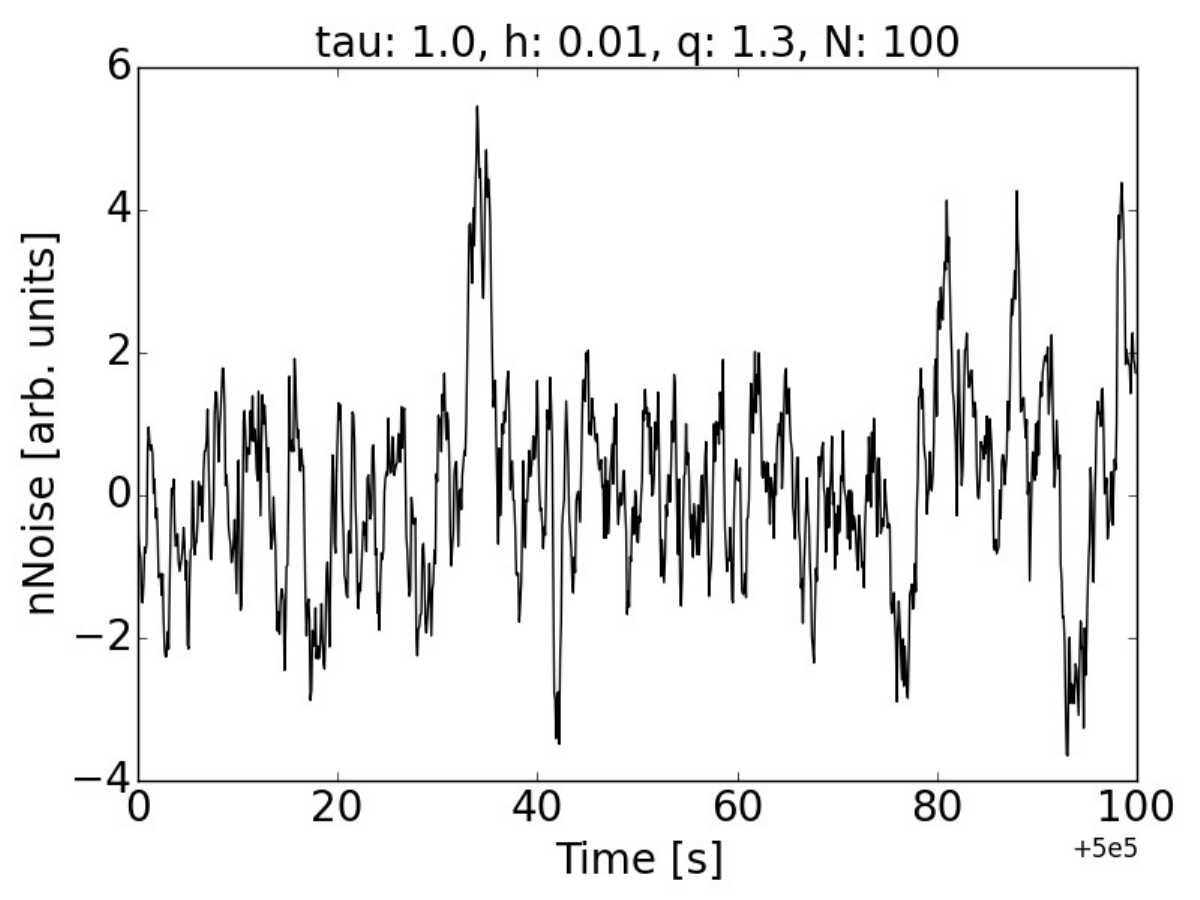}
\includegraphics[width=.45\textwidth]{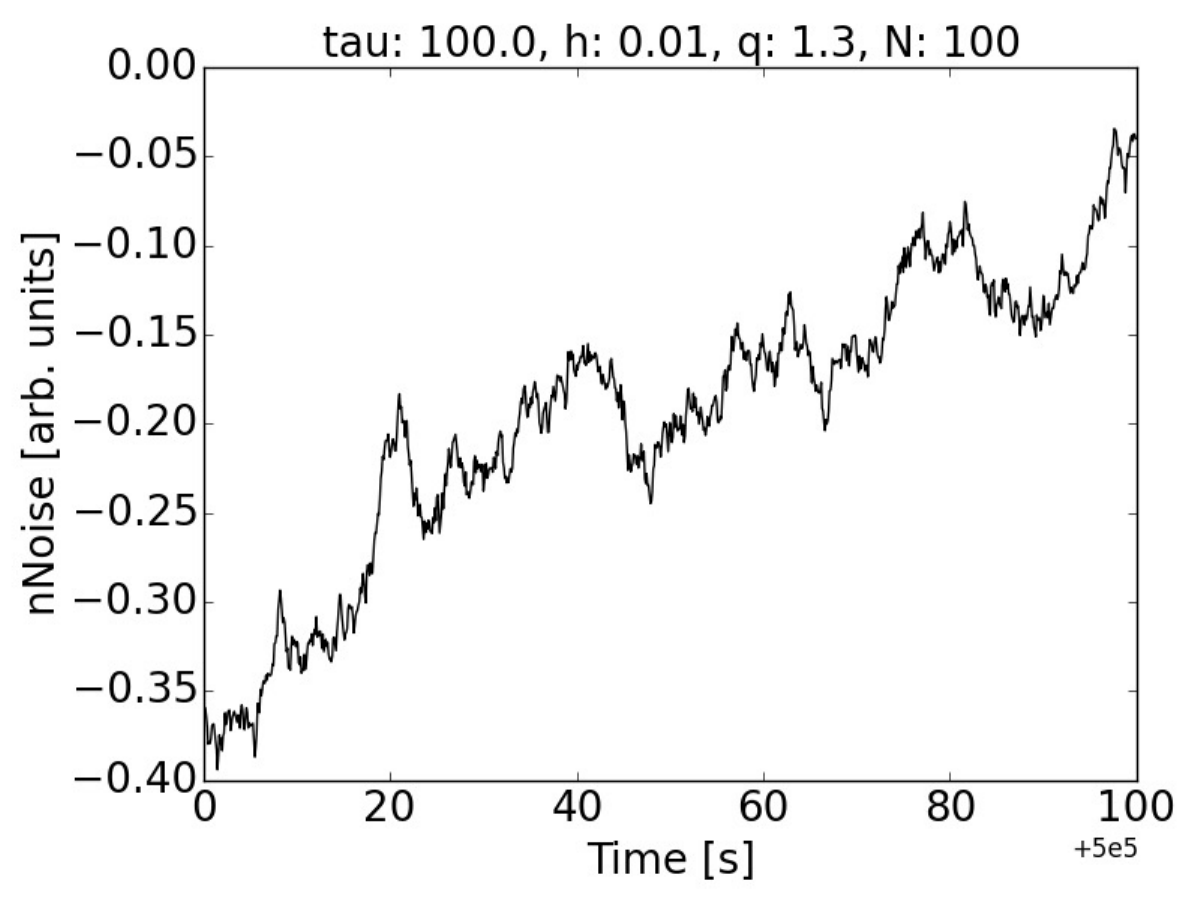}
\includegraphics[width=.45\textwidth]{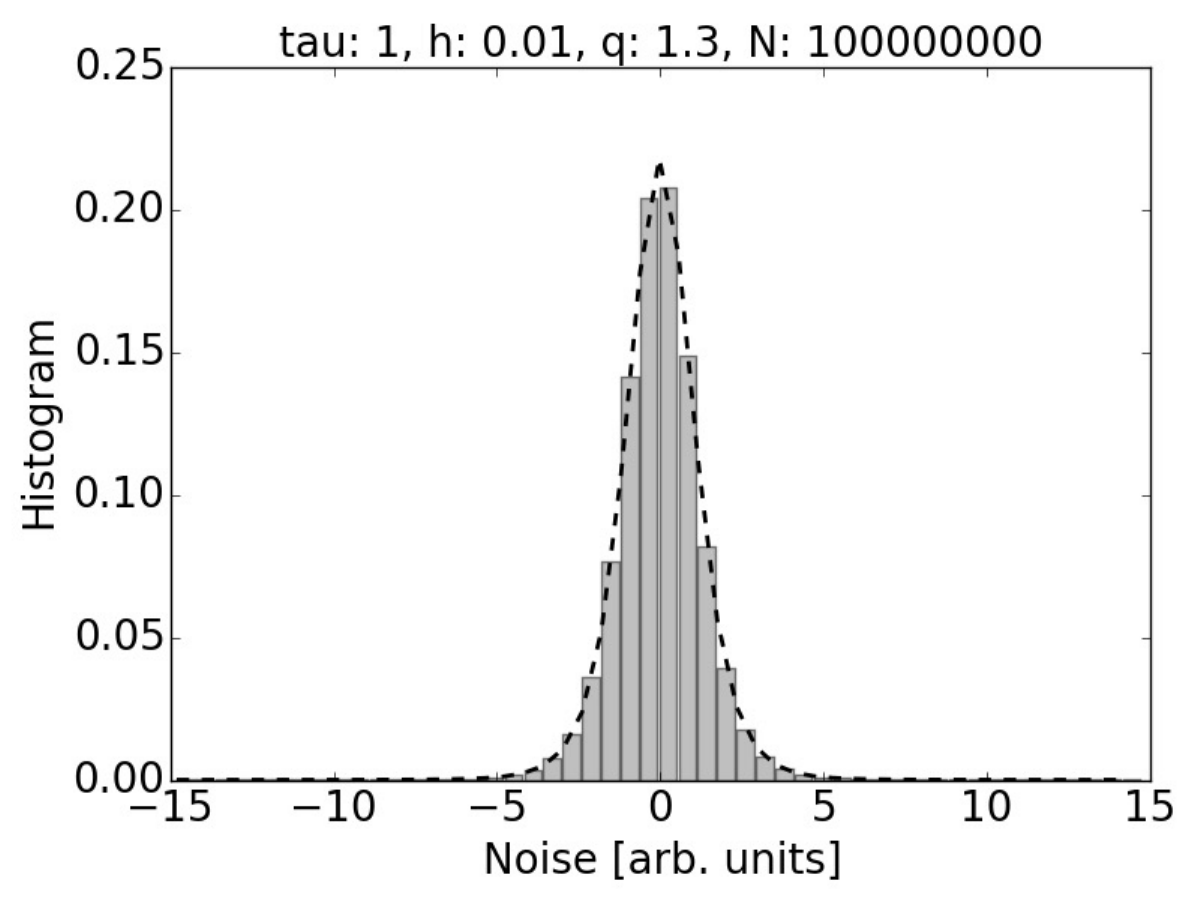}
\includegraphics[width=.45\textwidth]{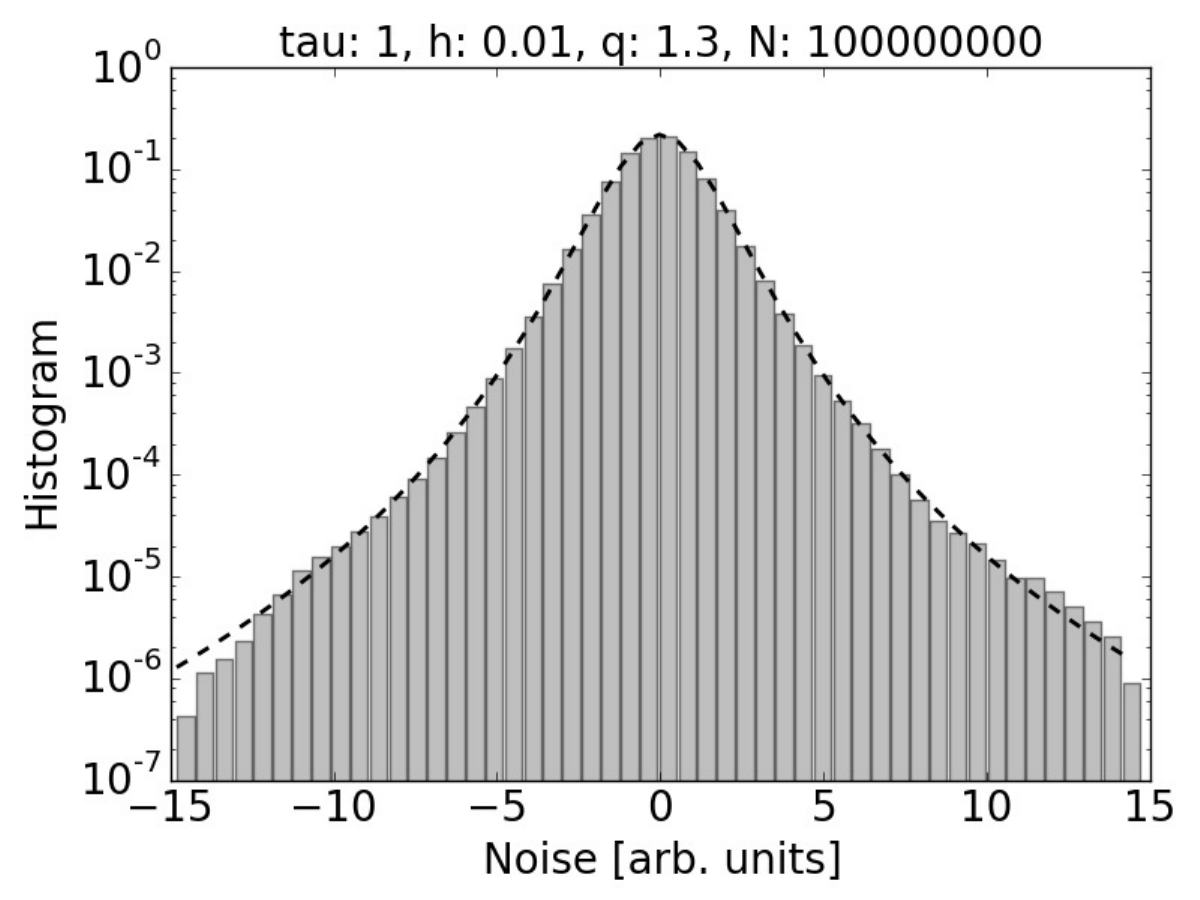}
\caption{q-noise for $q=1.3)$ (supra-Gaussian Behavior),  and integration step $h=0.01$. The top panels show a sample of the generated noise, for (left) $\tau=1 s$ and (right) $\tau=100 s$. Notice that in the right figure the noise is not centered around zero as it is performing a very long excursion (larger than the sample) given its very high autocorrelation. Both histograms in the bottom panels show the same data, concurring with the sample on top-left. ($\tau=1 s$). Although the linear histogram on the left, showing a bell-shaped distribution, could lead to suggest Gaussianity, the semi-logarithmic plot on the right, however, show a supra-Gaussian behavior. The dotted points curve shows the theoretical distribution as in Fig. \ref{fig:1} for the same parameters concurring perfectly with the histogram of the data. }\label{fig:qmorethan1}
\end{center}
\end{figure}

The noise presented here (Fig \ref{fig:qmorethan1}) is of a finite variance. The long excursions are however much longer and much more frequent than in the Gaussian case. This case is the most commonly used in the  applications of non-Gaussian noise presented below, as it allows to model many realistic systems outside of equilibrium .

%\clearpage
\section{Applications}\label{sec:5}

Some applications of the generated noise are shown below, both as an illustration and to show the potential of the generator both for research as well as a source of high quality noise for simulations or experiments (e.g. in electronics, optics, photosensitive chemical reactions etc\ldots). 

\subsection*{Stochastic resonance}

This is a phenomenon occurring in some nonlinear systems, whereby enhancing the response to a weak external signal may require \emph{increasing} the noise intensity.  An often resorted-to measure is the \emph{signal-to-noise ratio} at the input frequency $\omega$ (denoted by $R$).

The main numerical and theoretical results are \cite{futw01,ftwt03}: (1) for fixed $\tau$, the maximum $R$ increases with decreasing $q$; (2) for given $q$, the optimal noise intensity (the one maximizing $R$) decreases with $q$ and its value is approximately independent of $\tau$; (3) for fixed noise intensity, the optimal value of $q$ is independent of $\tau$ and in general turns out to be $q_\mathrm{op}\neq1$. A simple stochastic resonance experiment with a non-Gaussian \emph{white} noise \cite{ckfw01} confirmed most of these predictions.

\subsection*{Brownian motors}

A class of non-equilibrium systems with both potential technological applications and biological interest are the so called ``ratchets", in which the breakdown of spatial and/or temporal symmetry induces directional transport. Their transport properties can be studied by means of the Langevin equation
\begin{equation}\label{xpto}
m\frac{d^2 x}{dt^2}=-\gamma\frac{dx}{dt}-V'(x)-F+\xi(t)+\eta(t),
\end{equation}
with $m$ the particle's mass, $\gamma$ the friction constant, $V(x)$ the (sawtooth-like) ratchet potential, $F$ a constant ``load'' force, and $\xi(t)$ the thermal noise, satisfying $\langle\xi(t)\xi(t')\rangle=2\gamma T\delta(t-t')$. 

The system is kept out of thermal equilibrium by the time-correlated forcing $\eta(t)$ (with zero mean), allowing to rectify the motion. The $q$-dependence of the usual measures of performance has been studied: the mean current $J\equiv\langle dx/dt\rangle$ and the efficiency $\varepsilon$ (the ratio of the work per unit time done against $F$, to the mean power injected by $\eta$). 

In the overdamped regime $(m=0,\gamma=1)$, $J$ is found to grow monotonically with $q$ whereas $\varepsilon$ is maximized for some $1<q<5/3$. For $m\neq0$, ratchets exhibit mass-separation capabilities which are enhanced by non-Gaussian noise \cite{bowi04,bowi05}. In \cite{mawi08}, effects of biological and technological relevance have been found in a model for the transport properties of motor proteins when departing from Gaussian behavior: $J$ is maximized not only by an optimal noise intensity but also by an optimal $q\neq1$.

\subsection{Resonant gated trapping} \label{sec:resGate}

Stochastic resonance, which is essentially a threshold phenomenon, plays also a relevant role in ionic transport through cell membranes. In \cite{resw02}, a ``toy model'' considering the simultaneous action of a deterministic and a stochastic external field on the trapping rate of a gated imperfect trap, was studied by assuming Tsallis' noise with $q<1$: the bounded character of the PDF contributed positively to the rate of overcoming the threshold, and such rate remained at about the same order within a larger range of values than if $\eta$ had been a white noise.

\subsection*{Noise-induced transition}

A genetic model exhibiting a re-entrance from a disordered state to an ordered one, and  again to a disordered state as $\tau$ varies from $0$ to $\infty$ showed moreover a strong shift in the transition line, as $q$ departed from $q=1$. The transition was anticipated for $q>1$, while it was retarded for $q<1$ \cite{wito04}.

\subsection*{Noise-induced phase transition}

In fact fat-tail noise distributions ($q>1$) \emph{counteract} the effect of self-correlation (namely, they \emph{advance} the ordering boundary as $D$ is increased at constant coupling), and compact-support ones ($q<1$) \emph{enhance} it (they \emph{retard} the ordering boundary). Particular interest rises the effect of ($q<1$) multiplicative noise on the susceptibility: it shifts from being larger on the \emph{ordering} boundary to being larger on the \emph{disordering} boundary \cite{dewf07,defw07}.

An example of this phenomenon can be found in climate change. Many climatic ``Tipping points'' are are, in fact, noise-induced phase transitions whose forcing them (including astronomic, natural and antropogenic noise) are not necessarily Gaussian. An ``Early Warning''\cite{Lenton2011} system of tipping points should include simulation considering the non-gaussianity of the stochastic forcing.

\subsection*{Broad-spectrum energy harvesting}

In piezoelectric energy harvesting from noise, a system obeying a square-well potential can strongly profit from the large correlated excursion occurring for $q>1$ \cite{ddw12}. 

\section{Conclusions}
A lightweight software is presented that generates a class of non-Gaussian, colored noise. This noise can be handily generated during numerical experiments, or fed to experiments via an interface. The software, alongside documentation, is provided on the online repository \texttt{Github} including examples and unit test results, with an open source license.
Instances of noise-induced phenomena arising when the system is submitted to (colored and non-Gaussian) noise sources with Tsallis' q-statistics, as applications, have been briefly explored. The above discussed results show that non-Gaussian noise can significantly change the system's response in many noise-induced phenomena, as compared with the Gaussian case. Moreover, in all the cases presented here, the system's response was either enhanced or altered in a relevant way for values of $q$ departing from Gaussian behavior. In other words, the \emph{optimum response} occurred for $q\neq1$. Clearly, the study of the change in the response of other related noise-induced phenomena when subjected to such kind of non-Gaussian noise will be of great interest.
\section*{Acknowledgments}
The authors express their recognition to H.S. Wio and R. Toral, as sources of inspiration and fruitful conversations. \vskip.5cm

\bibliographystyle{elsarticle-num}
\bibliography{qNoise}






% \end{thebibliography}

\end{document}